\documentclass[aps,prl,twocolumn,groupedaddress,showpacs]{revtex4}
\usepackage{graphicx}
\draft
\begin{document}
\title{
Large Magnetoresistance Induced by Quantum Charge Fluctuations
in Magnetic Double Dots }
\author{L. Sheng$^{1}$, D. Y. Xing$^{1}$ and D. N. Sheng$^{2}$}
\address{$^1$National Laboratory of Solid State Microstructures, Nanjing University, Nanjing 210093, China\\
$^2$Department of Physics and Astronomy, California State
University, Northridge, CA 91330, USA}

\begin{abstract}
We study electron tunneling through two small ferromagnetic
dots. Quantum charge fluctuations and interdot
coupling make each Coulomb peak of the conductance at zero interdot
coupling split across. The interdot tunnel coupling is determined
by the relative orientation of magnetizations of the two dots,
leading to different splitting energies of the Coulomb peaks in parallel and
antiparallel magnetization alignments. As a result, a very large
tunneling magnetoresistance occurs near the Coulomb peaks, and its
sign may be either positive or negative.
\end{abstract}
\mbox{}\\
\pacs{PACS Number: 75.47.Jn, 73.23.Hk, 73.40.Gk, 75.45.+j}

%
\maketitle


The tunneling conductance of a tunnel junction made of two
ferromagnetic (FM) electrodes separated by a nonmagnetic (NM)
insulating layer may increase by several tens percent, as the
magnetizations of the FM electrodes change from antiparallel (AP)
to parallel (P) alignment under an external magnetic
field~\cite{TMR1,TMR2,TMR3,TMR4,TMR5,TMR6}. Such a
tunneling magnetoresistance (TMR) effect has important
applications in magnetic sensors, random access memories, and
magnetic imaging. The TMR originates from the spin polarization of
tunneling electrons in the FM electrodes.
The TMR ratio can be defined in a symmetric form as
\begin{equation}
\mbox{TMR}=\frac{G_{\mbox{\tiny P}}-G_{\mbox{\tiny AP}}}
{G_{\mbox{\tiny P}}+G_{\mbox{\tiny AP}}}\ ,\label{TMR}
\end{equation}
where $G_{\mbox{\tiny P}}$ and $G_{\mbox{\tiny AP}}$ are the
conductances for parallel and antiparallel magnetization
alignments, respectively. According to the simple
model~\cite{TMR1,TMRTheory1}, $\mbox{TMR}=P_{\mbox{\tiny
L}}P_{\mbox{\tiny R}}$ for the magnetic tunnel junctions, where
$P_{\mbox{\tiny L(R)}}$ is the spin polarization factor in the
left (right) FM electrode. For electrodes with spin polarization
of the same sign, the TMR ratio is usually positive. There is great
interest in inverse (negative)
TMR~\cite{InverseTMR1,InverseTMR2,InverseTMR3,InverseTMR4,InverseTMR5},
which is usually associated with certain hiding mechanisms that affect the
spin-dependent transport significantly. The inversion of TMR has been
attributed to the reversion of the spin polarization in either of two FM
electrodes due to the bonding effect at the FM/NM
interface~\cite{InverseTMR1,InverseTMR2,InverseTMR3}, and to the resonant
tunneling via localized states on impurities in the insulating
barrier~\cite{InverseTMR4,InverseTMR5}.

With the advancement of technology, magnetic tunnel junctions are
made increasingly small and controllable in recent years. Single
electron transistors consisting of FM metals~\cite{SET1} and
junctions with small FM islands~\cite{SET2} have been fabricated.
Presently, it is of particular interest to study the TMR through
small junctions and quantum dots, where the Coulomb interaction
plays an important
role~\cite{DotTMR2,DotTMR3,DotTMR4,DotTMR5,DotTMR6}. It has been
suggested that the TMR between two FM leads through an FM dot in
the Coulomb blockade regime may be enhanced by higher-order
electron tunneling via virtual intermediate states in the dot
(cotunneling)~\cite{DotTMR2,DotTMR3}, or oscillate in magnitude
with external voltage~\cite{DotTMR4,DotTMR5}. The TMR through
an NM dot may be unchanged or in some cases decreased by  Coulomb
blockade~\cite{DotTMR3,DotTMR6}.

Nonmagnetic quantum dots have previously
been extensively studied,
and many features of the tunneling conductance have been well
known~\cite{qDotReview}. The conductance of a quantum dot displays
Coulomb peaks at certain gate voltages, and is strongly suppressed
elsewhere by the Coulomb blockade~\cite{qDot1,qDot2,qDot3,qDot4}.
Electron transport through coupled double quantum dots has also
been extensively explored. In a double-dot structure, the quantum
charge fluctuations can be more directly probed than in a single
dot. As has been studied both experimentally~\cite{DoubleDotExp1}
and
theoretically~\cite{DoubleDotTheory1,DoubleDotTheory2,DoubleDotTheory3},
the interdot coupling and charge fluctuations make each Coulomb
peak at zero interdot couplings split across. However, spin-dependent
electron transport through coupled FM double quantum dots has not
yet been investigated so far. More importantly, we find that,
in the FM double quantum dots, a largest TMR effect may occur
essentially because of the quantum charge fluctuations between the
FM dots.

In this paper, we consider a magnetic tunnel junction consited of
two coupled FM quantum dots. The system is weakly coupled to two reservoirs of
electrons. A quantum-circuit description for the charge
fluctuations is developed. We show that, with changing the
magnetizations of the two FM dots from AP to P
alignment, the interdot tunnel coupling increases, which in turn
enhances the Coulomb peak splitting. As a consequence, the Coulomb
peaks shift to different gate voltages for AP and P alignments,
resulting in a very
large TMR. Moreover, depending on the gate voltage, the TMR ratio
may be either positive or negative.

The equivalent electrostatic circuit for the double-dot tunnel
junction is shown in Fig.\ 1. For simplicity, the two FM
dots are assumed identical. For the symmetric capacitors:
 $C_{\mbox{\tiny L}}=C_{\mbox{\tiny R}}=C_0$ and $C_{\mbox{\tiny
G}1}=C_{\mbox{\tiny G}2}=C_{\mbox{\tiny G}}$, we can confine
ourselves to the case of the same gate voltage $V_{\mbox{\tiny
G}1}=V_{\mbox{\tiny G}2}=V_{\mbox{\tiny G}}$. The qualitative
results of this work can be reproduced for nonidentical dots,
provided that the two gate voltages $V_{\mbox{\tiny G}1}$ and
$V_{\mbox{\tiny G}2}$ are allowed to change differently so that
electrons tunneling through the two dots resonates simultaneously.
Let $N_1$ and $N_2$ denote the numbers of excess electrons in
the two dots. The model Hamiltonian for the system can be written
in the form
\begin{equation}
H=K+U(N_1, N_2)\ ,\label{HAMIL}
\end{equation}
where the expression for electron kinetic energy $K$ will be
derived later, and the electrostatic energy of
the system is given by
\begin{equation}
U(N_1, N_2)=E_{\mbox{\tiny C}}(\delta^2N_{1}+\delta^2N_{2})+
2E'_{\mbox{\tiny C}}\delta N_{1}\delta N_{2}\ .\label{ELECTRO}
\end{equation}
Here $\delta N_{1}=N_{1}-N_{\mbox{\tiny C}}$ and $\delta
N_{2}=N_{2}-N_{\mbox{\tiny C}}$ with $N_{\mbox{\tiny
C}}=C_{\mbox{\tiny G}}V_{\mbox{\tiny G}}/e$. $E_{\mbox{\tiny
C}}=e^2C_{\mbox{\tiny $\Sigma$}}/2F$ and $E'_{\mbox{\tiny
C}}=e^2C_{\mbox{\tiny D}}/2F$ with $C_{\mbox{\tiny
$\Sigma$}}=C_{0} +C_{\mbox{\tiny G}}+C_{\mbox{\tiny D}}$ and
$F=C_{\mbox{\tiny $\Sigma$}}^2-C_{\mbox{\tiny D}}^2$.
\begin{figure}
\includegraphics[width=2.6in]{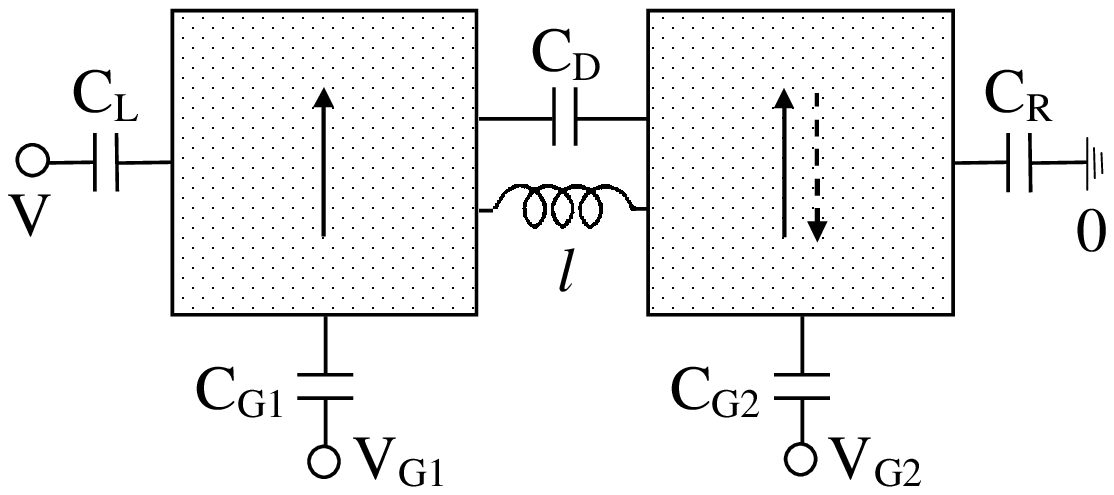}
\caption{Equivalent electrostatic circuit of a magnetic tunnel
junction consisting of two small ferromagnetic quantum dots, which
are weakly coupled to two electron reservoirs. The arrow
represents the magnetization direction of the FM dot. }
\end{figure}

The interdot tunnel coupling of electrons is assumed to be
much stronger than the tunnel coupling between the dots and the
reservoirs. As a result, for given total electron number ${\cal
N}=N_{1}+N_{2}$, quantum fluctuations of ${\cal M}=N_1-N_2$
are important. In previous treatments of the charge
fluctuations, very different assumptions were made on
the interdot tunnel matrix elements, e.g., from the
thin-shell model with one-to-one level
coupling~\cite{DoubleDotTheory1} to the thick-shell model with
single site-to-site
connection~\cite{DoubleDotTheory2,DoubleDotTheory3}. Here we
extend the approach developed by Girvin $et$ $al.$~\cite{Girvin}
to the charge fluctuations in a Coulomb-blockaded single junction
coupled to its environment by
long transmission lines. In the approach, the collective motion of
the charges is described by inductances, and the charge numbers
are treated as quantum canonical coordinates. In this essence, we
introduce an inductance $l$ between the dots to take the inter-dot
charge fluctuations into account. The advantage of the approach is
that it enables us to capture the essential physics of the charge
fluctuations without detailed assumptions about the tunnel matrix
elements. It will be shown that a simple connection
can be established between our theory and the previous
theories~\cite{DoubleDotTheory2,DoubleDotTheory3}.

Since inductance $l$ essentially describes the charge transfer
between the dots, it must depend on the magnetization alignment
of the FM dots because of the spin polarization of
electrons in the FM dots, just as in an ordinary FM/NM/FM tunnel
junction~\cite{TMR1,TMRTheory1}. For the P (AP)
alignment, the electron tunneling between the dots is relatively
easy (difficult), and so the inductance $l=l_{\mbox{\tiny P}}$
($l=l_{\mbox{\tiny AP}}$) is relatively small (large). According to
classical electromagnetism, as a current of $\dot{N}_1e=-\dot{N}_2e
=\dot{\cal M}e/2$ flows through inductance $l_{\eta}$
with $\eta=\mbox{P}$ or $\mbox{AP}$, the energy stored in the
inductance is $K=l_{\eta}e^2\dot{\cal M}^2/8$. Its corresponding
quantum form is $K=2P_{\mbox{\tiny ${\cal M}$}}^2/l_{\eta}e^2$
with $P_{\mbox{\tiny ${\cal M}$}}$ the momentum operator conjugate
to ${\cal M}$. Projecting $K$ into the subspace of integer
electron numbers $N_{1}$ and $N_{2}$~\cite{LCCircuit}, we obtain
$K=-(D_{\mbox{\tiny ${\cal M}$}}^\dagger +D_{\mbox{\tiny ${\cal
M}$}}-2)/2l_{\eta}e^2$, where operator $D_{\mbox{\tiny ${\cal
M}$}}^\dagger =e^{-2iP_{\mbox{\tiny ${\cal M}$}}/\hbar}$
increases ${\cal M}$ by two. Subsequently, $K$ is transformed into
the representation of $N_1$ and $N_2$, yielding
\begin{equation}
K=-\kappa_{\eta}(D_1^\dagger D_2+D_1D_2^\dagger)\ ,\label{KINETIC}
\end{equation}
where $\kappa_{\eta}=1/2l_{\eta}e^2$,
$D_{1(2)}^\dagger=e^{-iP_{1(2)}/\hbar}$ increases $N_{1(2)}$ by
unity with $P_{1(2)}$ the momentum conjugate to $N_{1(2)}$, and an
irrelevant constant energy $1 / l_{\eta}e^2$ has been neglected.
Equation (4) is a reasonable result, indicating the movement of an
electron from one dot to the other.

The above model is essentially a quantum LC circuit, whose general
eigenstates can be determined from the complicated Mathieu
equation~\cite{LCCircuit}. In this work, under the assumption of
$E_{\mbox{\tiny C}}\gg E'_{\mbox{\tiny C}}$ and $E_{\mbox{\tiny
C}}\gg\kappa_{\eta}$ and so to a good approximation, we
project the Hamiltonian into the subspace spanned by the
lowest-energy states for all possible ${\cal N}$. The eigenstates
of $N_1$ and $N_2$, $\vert
n_1,n_2\rangle=(D^\dagger_{1})^{N_1}(D^\dagger_{2})^{N_2} \vert
0,0\rangle$, can be used as the base wave vectors, where $\vert
0,0\rangle$ stands for the state without excess electrons in the
dots. For even ${\cal N}=2k$, the
low-energy state is the single state $\vert k,k\rangle$ with
electrostatic energy $E_{\mbox{\tiny ${\cal N}$}}=U(k,k)$; and for
odd ${\cal N}=2k + 1$, the low-energy
states are the two degenerate
states $\vert k+1, k\rangle$ and $\vert k, k+1\rangle$ with
electrostatic energy $E_{\mbox{\tiny ${\cal N}$}}=U(k+1, k)$.
Since the kinetic energy given by Eq.\ (\ref{KINETIC}) conserves the
total electron number ${\cal N}$, the only nonzero matrix elements of
$K$ in the low-energy subspace are those between the two
degenerate states for odd ${\cal N}=2k+1$, namely, $\langle
k+1,k\vert K\vert k,k+1\rangle= \langle k,k+1\vert K\vert
k+1,k\rangle=-\kappa_{\eta}$. It is thus easy to obtain the
eigenstates of the system. For an even electron number (${\cal
N}=2k$), the eigenstate is $\vert {\cal N}\rangle =\vert
k,k\rangle$ with eigenenergy ${\cal E}_{\mbox{\tiny ${\cal N}$}}
=E_{\mbox{\tiny ${\cal N}$}}$. For an odd electron number (${\cal
N}=2k+1$), there are two eigenstates, which will be distinguished
from each other by symbols $\uparrow$ and $\downarrow$,
symmetric state $\vert {\cal N}\uparrow\rangle= (\vert k+1,
k\rangle+\vert k, k+1\rangle)/\sqrt{2}$ with lower eigenenergy
${\cal E}_{\mbox{\tiny ${\cal N}$}\uparrow}= E_{\mbox{\tiny ${\cal
N}$}}-\kappa_{\eta}$ and antisymmetric state $\vert {\cal
N}\downarrow\rangle= (\vert k+1, k\rangle-\vert k,
k+1\rangle)/\sqrt{2}$ with higher eigenenergy ${\cal
E}_{\mbox{\tiny ${\cal N}$}\downarrow}=E_{\mbox{\tiny ${\cal
N}$}}+\kappa_{\eta}$.

The total excess electron number ${\cal N}$ in the ground state is
determined by the minimum of system eigenenergy. Therefore,
tunneling of an electron into or out of the dots usually causes an
increase of the system energy and so the tunneling rate is
suppressed, leading to the Coulomb blockade effect.
At certain values of the gate voltage, however, the system
eigenenergies are degenerate for different ${\cal N}$, i.e.,
$E_{2k}=E_{(2k+1)\uparrow}$ or $E_{2k+2}=E_{(2k+1)\uparrow}$,
in which there appear Coulomb peaks of linear conductance. The
positions of the Coulomb peaks are obtained as
\begin{equation}
V_{\mbox{\tiny G}}=V_{\mbox{\tiny G},{\eta}}^{\pm}\equiv\frac{e}{C_{\mbox{\tiny G}}}
\left(k+\frac{1}{2}\pm\frac{\delta_{\eta}}{2}\right)\ ,\label{RESONANCE}
\end{equation}
where $\delta_{\eta}=(E'_{\mbox{\tiny
C}}+\kappa_{\eta})/E_{\mbox{\tiny C}}$ describes the correction to
the peak position due to interdot electrostatic coupling
$E'_{\mbox{\tiny C}}$ and interdot tunnel coupling $\kappa_{\eta}$.
According to Eq.\ (\ref{RESONANCE}), if the interdot coupling is
absent, the conductance peaks occur periodically, whenever the gate
voltage is equal to half-integer times $e/C_{\mbox{\tiny G}}$. The
presence of the interdot coupling splits each peak into two, the
distance between them given by
$\delta_{\eta}e/C_{\mbox{\tiny G}}$. It is interesting to notice
that our result recovers the previous result for the NM dots obtained
in the weak-tunneling regime~\cite{DoubleDotTheory2,DoubleDotTheory3},
if we set
$\kappa_{\eta}=(\hbar\ln 2/\pi) G_{\mbox{\tiny D}}/C_{\mbox{\tiny
$\Sigma$}}$ with $G_{\mbox{\tiny D}}$ the interdot conductance.
It then follows that, for the FM dots, $\kappa_{\eta}$ does depend
on the relative orientation of their magnetizations, for tunneling
conductance
$G_{\mbox{\tiny D}}$ between two FM dots must be orientation dependent.

According to Eq.\ (\ref{RESONANCE}), there is a position shift of the
Coulomb peaks with changing the magnetization alignment from AP to P.
This effect originates from the quantum charge fluctuations and
results in a very large TMR. To evaluate the conductance and TMR,
we will confine ourselves to the vicinities of the Coulomb peaks,
where real tunneling processes of the electrons dominate. In
the strongly Coulomb-blockaded region, higher-order tunneling
processes of electrons via virtual intermediate states
(cotunneling) may be important~\cite{DotTMR2,DotTMR3}, which will not
be considered here. Let us consider the first pair of
resonant peaks for positive gate voltage $V_{\mbox{\tiny G}}$,
whose positions are determined from Eq.\ (\ref{RESONANCE}) by
setting $k=0$. The physics around other peaks is completely similar.
The conductance is given by~\cite{DoubleDotTheory3}
\begin{equation}
G_{\eta}=\frac{G_0}{2Z}\left[g(V_{\mbox{\tiny G}}-V^-_{\mbox{\tiny
G},\eta})+g(V^+_{\mbox{\tiny G},\eta}-V_{\mbox{\tiny G}})\right]\
,
\end{equation}
where $G_0$ is the conductance of the dot-reservoir barrier,
$Z=2+e^{-\alpha(V_{\mbox{\tiny G}}-V^-_{\mbox{\tiny
G},\eta})}+e^{-\alpha(V^+_{\mbox{\tiny G},\eta}-V_{\mbox{\tiny
G}})}$ is the partition function with $\alpha=2C_{\mbox{\tiny
G}}E_{\mbox{\tiny C}}/ek_{\mbox{\tiny B}}T$,
and $g(x)=\alpha x/(e^{\alpha x}-1)$.

The calculated conductances for P and AP
magnetization alignments as functions of normalized gate voltage
for several temperatures are shown in Figs.\ 2a and 2b,
respectively. As expected, the position of the resonant peaks
depends on the magnetization alignment. The distance between the
split peaks is greater in the P alignment than in the
AP alignment, the resulting TMR ratio plotted in Fig.\ 2c.
On either side of the two peaks for $G_{\mbox{\tiny AP}}$, the TMR
is positive, exhibiting the same sign as that in normal FM/NM/FM
junctions. At low temperatures, e.g.,
$k_{\mbox{\tiny B}}T=0.01E_{\mbox{\tiny C}}$, the positive
TMR may reach about $98\%$, corresponding to $G_{\mbox{\tiny
P}}/G_{\mbox{\tiny AP}}\simeq 100$. It forms a striking contrast to
the much smaller ratio $\kappa_{\mbox{\tiny
P}}/\kappa_{\mbox{\tiny AP}}=2$ used in the calculation. Actually,
for gate voltage $V_{\mbox{\tiny G}}$ on either side of the two peaks
for $G_{\mbox{\tiny P}}$, one can obtain
 $G_{\mbox{\tiny P}}/G_{\mbox{\tiny AP}}\simeq\exp[
(\kappa_{\mbox{\tiny P}}-\kappa_{\mbox{\tiny AP}})/k_{\mbox{\tiny B}}T]$,
increasing exponentially with the temperature lowered.
More interestingly, in the region between the two peaks of
$G_{\mbox{\tiny AP}}$, the TMR  is not only very large in magnitude
but also inverted in sign. For gate voltage in this region, the
conductance ratio $G_{\mbox{\tiny P}}/G_{\mbox{\tiny AP}}\simeq
\exp[-(\kappa_{\mbox{\tiny P}}-\kappa_{\mbox{\tiny
AP}})/k_{\mbox{\tiny B}}T]$ decreases exponentially with
the temperature lowered. With
increasing temperature, both positive TMR and negative TMR
decrease in magnitude. As energy $k_{\mbox{\tiny B}}T$ is increased
so as to be comparable with the energy
separation $(\kappa_{\mbox{\tiny P}} -\kappa_{\mbox{\tiny AP}})$
of the left (or right) two Coulomb peaks for $G_{\mbox{\tiny AP}}$ and
$G_{\mbox{\tiny P}}$, the TMR ratio becomes very small.
\begin{figure}
\includegraphics[width=2.6in]{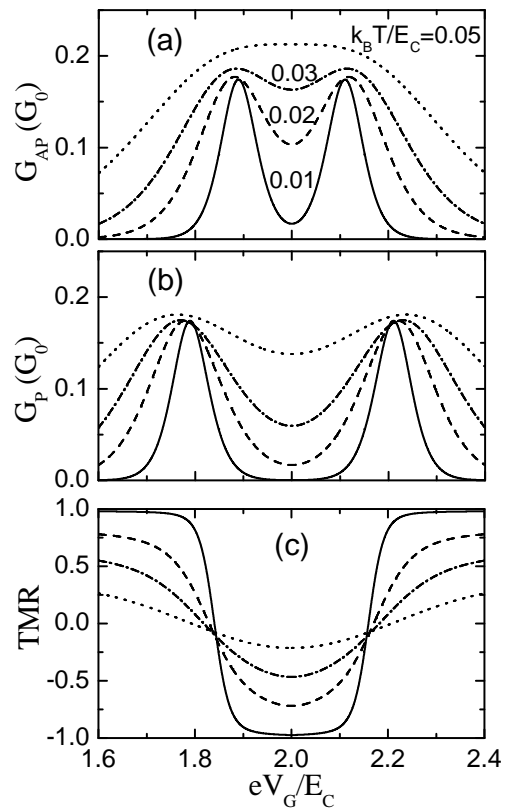}
\caption{Conductances for antiparallel magnetization alignment (a)
and parallel magnetization alignment (b), and TMR ratio as a
function of normalized gate voltage $eV_{\mbox{\tiny
G}}/E_{\mbox{\tiny C}}$ (c). Here $C_{\mbox{\tiny D}}=0$,
$C_{0}=C_{\mbox{\tiny G}}$, and $\kappa_{\mbox{\tiny
P}}=2\kappa_{\mbox{\tiny AP}}=0.1E_{\mbox{\tiny C}}$.}
\end{figure}

In Fig.\ 3, the conductance ratio $G_{\mbox{\tiny
P}}/G_{\mbox{\tiny AP}}$ is plotted as a function of temperature
for different interdot electrostatic couplings $C_{\mbox{\tiny
D}}$ and interdot tunnel coupling $\kappa_{\eta}$. We note that
$G_{\mbox{\tiny P}}/G_{\mbox{\tiny AP}}>1$ corresponds to positive
TMR, and $G_{\mbox{\tiny P}}/G_{\mbox{\tiny AP}}<1$ corresponds to
negative TMR. So far, we have confined ourselves to the simple
case of zero interdot capacitor $C_{\mbox{\tiny D}}$. The interdot
capacitor represents the effect of the excess charges in one dot
on the electrical potential of the other dot. According to Eq.\
(\ref{RESONANCE}), a nonzero $C_{\mbox{\tiny D}}$ mainly enhances the
splitting of the Coulomb peaks. The two dotted lines in Fig.\ 3
are obtained for nonzero $C_{\mbox{\tiny D}}$. By comparing them
with the solid curves, we find that an increased interdot capacitance
does not change the TMR significantly.
\begin{figure}
\includegraphics[width=2.6in]{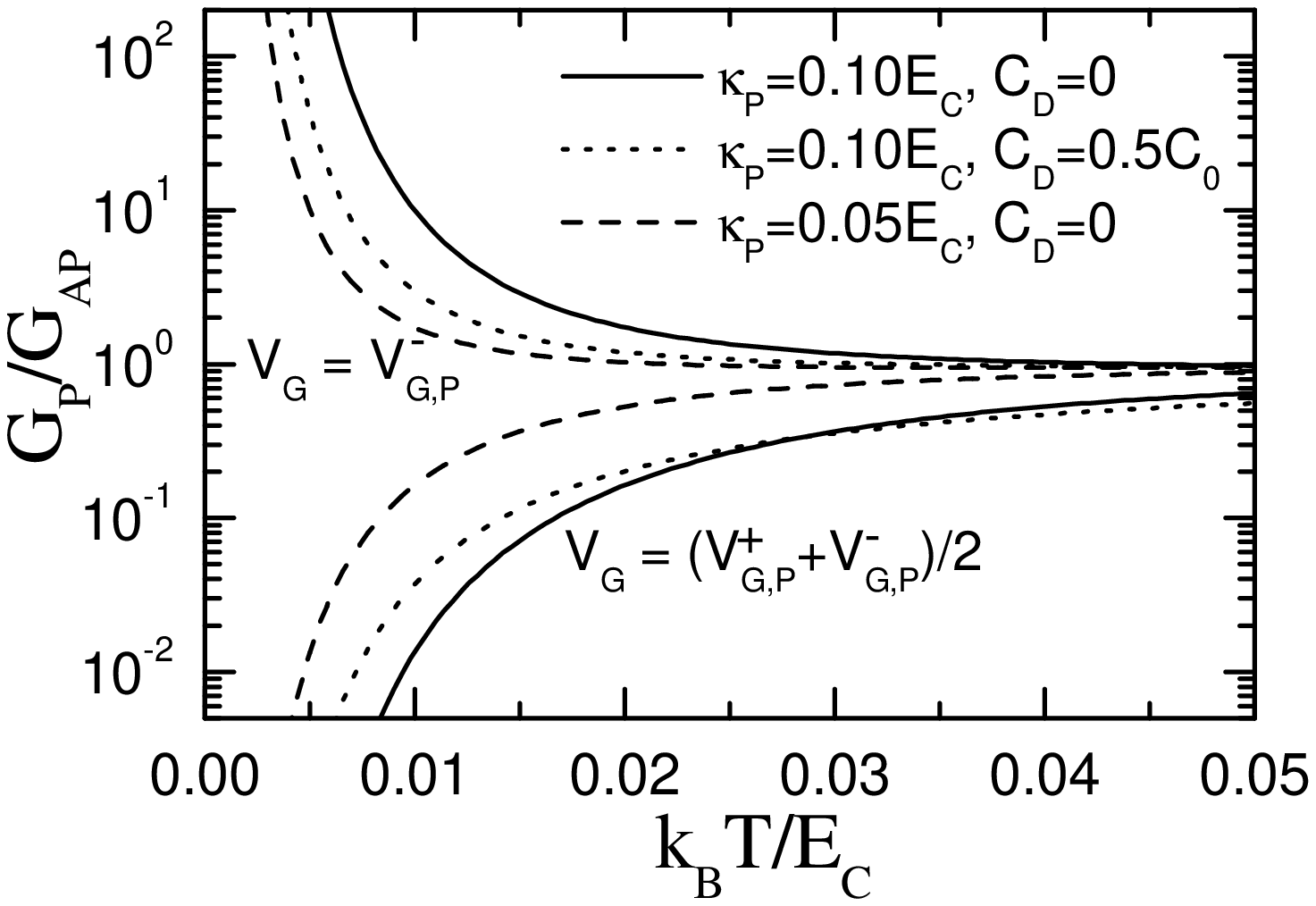}
\caption{Conductance ratio $G_{\mbox{\tiny P}}/G_{\mbox{\tiny
AP}}$ as a function of normalized temperature $k_{\mbox{\tiny
B}}T/E_{\mbox{\tiny C}}$ for different $\kappa_{\mbox{\tiny P}}$
and $C_{\mbox{\tiny D}}$. Here $C_{0}=C_{\mbox{\tiny G}}$ and
$\kappa_{\mbox{\tiny P}}=2\kappa_{\mbox{\tiny AP}}$. The gate
voltage is $V_{\mbox{\tiny G}}=V_{\mbox{\tiny G,P}}^{-}$ for the
upper three curves, and $V_{\mbox{\tiny G}}=(V_{\mbox{\tiny
G,P}}^{+} +V_{\mbox{\tiny G,P}}^{-})/2$ for the lower three
curves. }
\end{figure}

Finally, we wish to make a discussion on experimental
realization of the present theory. With today's fabrication technology,
it is possible to connect two FM dots by using short narrow
conductor (quantum wire)~\cite{QDOT_FAB}, and keep the FM dots sufficiently
separated in space. In this manner, relatively strong
electron hopping coupling can be ensured,
and in the meantime magnetic dipolar interaction can be avoided
between the FM dots. As a result, the RKKY interaction caused by
electron hopping between the dots may induce an AF
alignment, if the dots are suitably distanced.
The P alignment can be achieved simply by applying an external magnetic field.
As a simplified model of the real systems, we have omitted the complexities
such as spin-flip scattering and electron virtual tunneling in the Coulomb
blockade regime. These effects could reduce the TMR measured experimentally.
However, it is believed that they will not change the results qualitatively.

In summary, we have investigated the charge fluctuations and
conductance peak splitting in magnetic double quantum dots. The
characteristic energy of the Coulomb peak splitting is shown to
depend on the magnetization alignment. Therefore, the change in
the magnetization alignment causes very large TMR
effect. The TMR ratio may be either positive or negative, depending
on the value of gate voltage.

D. Y. X. thanks the support from the National Natural Science
Foundation of China under Grant No. 10374046 and 10174011.
D. N. S was supported by ACS-PRF \# 36965-AC5,
Research Corporation Grant CC5643, the NSF grants DMR-00116566
and DMR-0307170, and the KITP at Santa Barbara
through PHY99-07949.

\end{document}